\documentstyle[12pt]{article}

\begin{document}
\baselineskip=24pt

\title{
\vspace{-1.5cm}
\begin{flushright}
{\normalsize KEK-CP-14}\\
\vspace{-0.3cm}
{\normalsize KEK Preprint 94-11}\\
\vspace{-0.3cm}
{\normalsize April 1994 }\\
\end{flushright}
\vspace*{1.0cm}
{\Large An Exploratory Study of Nucleon-Nucleon Scattering Lengths \\
in Lattice QCD \\}
}

\author{ M. Fukugita$^{a)}$,  Y. Kuramashi$^{b)}$, H. Mino$^{c)}$,
         M. Okawa$^{b)}$, A. Ukawa$^{d)}$ \\ \\
$ ^a${\it Yukawa Institute, Kyoto University}\\
     {\it Kyoto 606, Japan}\\ \\
$ ^b${\it National Laboratory for High Energy Physics(KEK)}\\
     {\it Tsukuba, Ibaraki 305, Japan}\\ \\
$ ^c${\it Faculty of Engineering, Yamanashi University}\\
     {\it Kofu 404, Japan}\\ \\
$ ^d${\it Institute of Physics,  University of Tsukuba,Tsukuba}\\
     {\it Ibaraki 305, Japan}\\ \\
}

\date{}

\maketitle
\begin{abstract}
\baselineskip=24pt

An exploratory study is made of the nucleon-nucleon $s$-wave scattering
lengths
in quenched lattice QCD with the Wilson quark action.  The
$\pi$-$N$ and $\pi$-$\pi$ scattering lengths are also calculated for
comparison.  The calculations are made with  heavy quarks corresponding to
$m_\pi/m_\rho\approx 0.73-0.95$.  The results show that the $N$-$N$ system has
an attractive force in both spin-singlet and triplet channels, with their
scattering lengths significantly larger than those for the $\pi$-$N$ and
$\pi$-$\pi$ cases, a trend which is qualitatively consistent with the
experiment.  Problems toward a more realistic calculation for light quarks are
discussed.

\end{abstract}


\newpage
The understanding of low energy nucleon-nucleon$(N$-$N)$ interactions has been
one of the most fundamental problems in nuclear physics.  A large effort  to
construct elaborate meson exchange models has enabled us to  describe the
low energy $N$-$N$ phase shifts and deuteron properties quite
well\cite{potential}. However, these models are essentially phenomenological
in
that they use mesons and baryons as the basic degrees of freedom and choose
the
coupling strengths among them to fit the experimental data.  Deriving the low
energy properties of the $N$-$N$ interaction directly from the dynamics of
quarks and gluons is clearly a problem posed to numerical studies of lattice
QCD.

An important parameter characterizing the low energy $N$-$N$
interaction is the $s$-wave scattering length.  The
experimental values are listed in Table~1\cite{experiment}
together with those for  $\pi$-$N$ and $\pi$-$\pi$ cases for
comparison.  We observe that the $N$-$N$ scattering lengths
are markedly larger.  We also note that the negative sign for the
spin triplet channel reflects the presence of the deuteron bound
state as follows from Levinson's theorem.  The small values
for the  $\pi$-$N$ and $\pi$-$\pi$ cases are understood as a
consequence of chiral symmetry for $u$ and $d$ quarks;
current algebra and PCAC have successfully predicted the sign and
magnitude\cite{weinberg}.  Since no
such constraint exists for $N$-$N$ scattering, its large
scattering lengths are dynamical phenomena of QCD.

An elegant method for calculating scattering lengths in lattice QCD is
provided by the formula\cite{luescher}
relating the $s$-wave scattering length $a_0$ between two hadrons $h_1$ and
$h_2$ to the energy shift $\delta E=E_{h_1h_2}-(m_{h_1}+m_{h_2})$ of the two
hadron
state at zero relative momentum  confined in a finite
spatial box of a size $L^3$.  It is  given by
\begin{equation}\delta E=-\frac{2\pi
a_0}{\mu L^{3}}(1+c_{1}\frac{a_0}{L}
+c_{2}(\frac{a_0}{L})^{2})+O(L^{-6}) \label{eq:Lusc}  \end{equation}
with $\mu=m_{h_1}m_{h_2}/(m_{h_1}+m_{h_2})$ and
$c_{1}=-2.837297,c_{2}=6.375183$.  The energy shift and hadron masses
$m_{h_1}$
and $m_{h_2}$  can be extracted from hadron four- and two-point functions
calculated through numerical simulations.  Previous studies have shown the
practical feasibility of the application of the formula for the case of
$\pi$-$\pi$ scattering both for $I=2$\cite{earlywork} and also for the
technically more difficult case of $I=0$\cite{kuramashi}.  With the
development
of the technique to calculate hadron four-point functions\cite{kuramashi}, we
have attempted an exploratory  study of the $N$-$N$ $s$-wave scattering
lengths, which we report in this article.

A realistic calculation of the $N$-$N$ scattering lengths faces three basic
obstacles.  The large values of scattering
lengths of order 10fm to be obtained mean that lattice sizes much
larger than $2-3$fm, which are accessible in current numerical simulations,
will
be needed to suppress $O(L^{-6})$ corrections neglected in (\ref{eq:Lusc}).  A
further difficulty in the spin triplet channel is that the lowest scattering
state orthogonal to the bound deuteron state has to be constructed to apply
(\ref{eq:Lusc}).  Finally statistical fluctuations in $N$-$N$
four-point functions grow rapidly toward large time separations and small
quark
masses as we shall discuss in detail below.  In order to avoid these problems
in our initial study, we have carried out  simulations for heavy quarks with
$m_\pi/m_\rho\approx 0.73-0.95$ in the quenched approximation.  We
employ  the Wilson quark action and work at the inverse coupling constant
$\beta=6/g^2=5.7$ which corresponds to the lattice spacing $a\approx
0.14$fm.  A calculation of $\pi$-$N$ and $\pi$-$\pi$ scattering lengths at the
same $\beta$ is also carried out to compare them with the results for the
$N$-$N$ case.

The parameters of runs are tabulated in Table~2.  Gauge configurations are
generated for the single plaquette action separated by 1000 sweeps of the
pseudo-heat bath algorithm.  Calculations
for the $\pi$-$\pi$ and $\pi$-$N$ cases are made on a $12^3\times 20$ lattice,
while for the $N$-$N$ case we employ a $20^4$ lattice anticipating larger
scattering lengths.   We extract the energy shift $\delta
E=E_{h_1h_2}-(m_{h_1}+m_{h_2})$ from the ratio of the hadron Green's
functions,
\begin{eqnarray}
R(t)&=&{<{\cal O}_{h_1}(t) {\cal O}_{h_2}(t+1)
{\cal O}_{h_1}(0) {\cal O}_{h_2}(1)>
\over
<{\cal O}_{h_1}(t) {\cal O}_{h_1}(0)>
<{\cal O}_{h_2}(t+1) {\cal O}_{h_2}(1)>}\\ \nonumber
&\stackrel{t\rightarrow\infty}{\rightarrow}&
\exp (-\delta E t)
\label{eq:ratio}
\end{eqnarray}
where the operators for $h_2$ are shifted by one lattice unit in the
temporal direction to avoid mixing of color-Fierz transformed
contributions\cite{kuramashi}.

Quark propagators for the $N$-$N$ case are calculated with
a wall source at $t=0$ or 1, fixing gauge configurations to the
Coulomb gauge in all space-time.  Projecting
out spin singlet and triplet combinations of the two nucleon system is made by
non-relativistically  combining the upper components of the nucleon operator
$N=({}^tqC^{-1}\gamma_5q)q$ taking account of isospin factors.

Calculation of the $\pi$-$N$ and $\pi$-$\pi$ four-point functions requires
quark
propagators connecting two arbitrary space-time sites. We handle this problem
by the method proposed in our previous work\cite{kuramashi}: quark
propagators are calculated  with a wall source for every time slice without
gauge fixing. The nucleon source operator for the $\pi$-$N$ case, however, is
fixed to the  Coulomb gauge to increase signal to noise ratio.

In Fig.~1 we show $R(t)$ for the spin singlet and triplet channels in the
$N$-$N$
case at $K=0.160$ corresponding to $m_\pi/m_\rho=0.85$.  A clear signal with
a positive slope is observed for both channels, which means attraction
($\delta E_{NN}=E_{NN}-2m_N<0$) as expected.  Similar results are obtained at
two
other values of the hopping parameter $K=0.15(m_\pi/m_\rho=0.95)$ and
$0.164(m_\pi/m_\rho=0.73)$.  We extract the energy shift $\delta E_{NN}$ by
fitting $R(t)$ to a linear form $R(t)=Z(1-\delta E_{NN}t)$. The fitting range
is chosen to be $4\leq t\leq 9$ for $K=0.150$ and $0.160$. For the case of
$K=0.164$  we used $2\leq t\leq 6$ for the fit due to the poor quality of our
data.    The fitted values of $\delta E_{NN}$ are quite small $(\approx
0.01$),
justifying the use of a linear function instead of an exponential.

For heavy quark the $N$-$N$ interaction becomes shorter ranged since pions
exchanged between the nucleons are heavy, while the
size of the nucleon will not be much reduced.  In a simple potential model
with
a linear confining potential, for example, the hadron size scales as a third
power of the constituent quark mass which is a slowly varying function of the
hopping parameter.  We may expect then that the deuteron becomes unbound for
heavy quarks.  Assuming this to be the case we can extract the scattering
lengths through (\ref{eq:Lusc}) for both spin singlet and triplet channels.
The
results in lattice units are tabulated in Table~3.

Our results for the $\pi$-$N$ and $\pi$-$\pi$ four-point functions are plotted
in Fig.~2 for $K=0.164$.  The data are also fitted with the linear form
$R(t)=Z(1-\delta Et)$ over $4\leq t\leq 9$.  The
resulting scattering lengths are listed in Table~3 (we could not
extract the $I=1/2$ $\pi$-$N$ scattering length at $K=0.1665$
due to large errors in our data; this point will be discussed below). The
predictions of current algebra and PCAC are given by \begin{equation}
a_0(\pi\pi)=+{7\over 16\pi}{\mu_{\pi\pi}\over f_\pi^2}\  (I=0),\quad -{1\over
8\pi}{\mu_{\pi\pi}\over f_\pi^2}\  (I=2), \end{equation}
and
\begin{equation}
a_0(\pi N)=+{1\over 4\pi}{\mu_{\pi N}\over f_\pi^2}\  (I=\frac{1}{2}),\quad
-{1\over 8\pi}{\mu_{\pi N}\over f_\pi^2}\  (I=\frac{3}{2}),
\end{equation}
$\mu_{\pi\pi}=m_\pi/2$ and $\mu_{\pi N}=m_\pi m_N/(m_\pi +m_N)$ being the
reduced masses.  We list in Table~3 the numerical values of these
predictions for the values of $m_\pi, m_N$ and $f_\pi$ measured in our
simulations with the tadpole-improved renormalization factor for
$f_\pi$\cite{lepagemackenzie}.  Our results for the scattering lengths are
consistent with these values within one standard
deviation.  It is somewhat surprising that the agreement holds even at quite
heavy quarks corresponding to $m_\pi/m_\rho\approx 0.73$. ( For the
$\pi$-$\pi$
scattering lengths a
similar agreement has previously been reported\cite{earlywork,kuramashi}.)

In Fig.~3 we plot the scattering lengths as a
function of $m_\pi/m_\rho$.  It is apparent that the
$N$-$N$ scattering lengths are substantially larger than the  $\pi$-$N$
and $\pi$-$\pi$ scattering lengths already for a heavy quark
corresponding to $m_\pi/m_\rho\approx 0.73$.  Also noteworthy is the trend,
albeit with sizable errors, that the values for the spin triplet channel are
 larger than those for the singlet channel, indicating a stronger
attraction in the triplet channel, which is consistent with the existence of
the
deuteron bound state.  We find these results to be
encouraging, although the scattering lengths we obtained are still much small
compared to the experiment: our results
correspond to $a_0(NN)\approx 1.0-1.5$fm in physical units  with $a\approx
0.14$fm determined from the $\rho$ meson mass extrapolated to the chiral
limit.  These small $N$-$N$ scattering lengths are most likely to arise from
a
short interaction range for heavy pions ($m_\pi\approx 0.7-1.5$GeV) for which
our simulations are made; we expect that they increase if simulations would be
made with a small quark mass\cite{footnote}.

Let us now discuss the issues for extension of the present work toward a more
realistic case of light quarks.  On the theoretical side, if deuteron is a
bound
state only for light enough quarks as assumed here, the scattering length for
the triplet channel extracted from the lowest two-nucleon energy should
diverge
at the value of $K$ corresponding to the bound state formation, beyond which
we
need to construct the lowest scattering state orthogonal to the deuteron to
extract the scattering lengths.  Observing the diverging trend would be an
interesting problem in its own light.

From the view point of simulations reducing statistical errors presents a
major problem toward light quarks.  A simple argument indicates that the error
of
$R(t)$ grows exponentially with $t$ as $\delta R(t)\propto\exp (
(2m_N-3m_\pi)t )$ for the $N$-$N$ four-point function; since the pion mass
vanishes in the chiral limit while the nucleon mass stays non-zero, the slope
$\alpha_{NN}=2m_N-3m_\pi$ becomes larger for lighter quarks.  A rapid growth
of
error is clearly seen in Fig.~1; its $t$ dependence is roughly consistent
with the estimate.    We also expect that the magnitude of
errors increases for lighter quarks.   Thus a progressively larger
statistics is needed to obtain reliable results as the quark mass is
decreased.

A similar situation holds for the
$\pi$-$N$ case.  The estimated slope $\alpha_{\pi
N}=m_N-m_\pi/2$ is even larger than $\alpha_{NN}$ for the range
$K=0.1665-0.15$
where we have made our simulation ({\it e.g.,} $\alpha_{\pi N}=0.84$ and
$\alpha_{NN}=0.66$ for $K=0.164$).  This explains a faster increase of
errors with $t$ for the $\pi$-$N$ case (compare  Fig.~1 and
Fig.~2(a)).  An increase of magnitude of errors is also observed
as the quark mass is reduced from $K=0.164$ to $K=0.1665$.  This is the
reason why we could not obtain the $I=1/2$ $\pi$-$N$ scattering length for
$K=0.1665$.

To summarize, we have explored the possibility of extracting the $N$-$N$
scattering lengths from numerical simulations of lattice QCD.  Albeit carried
out at heavy quarks, our results show that the attractive forces work in
both channels of $s$-wave $N$-$N$ scattering,  its scattering lengths
being significantly larger compared to the $\pi$-$N$ and $\pi$-$\pi$
scattering
lengths.  It remains an important and realistic problem to confirm that
reducing
the quark mass results in an increase of scattering lengths to the values
observed experimentally.  Such a calculation, however, requires a much larger
lattice and smaller quark masses than are possible at present, and hence has
to
await a development of the computing power.

Numerical calculations for the present
work have  been carried out on HITAC S820/80 at KEK.  This work is supported
in
part by the Grants-in-Aid of the Ministry of Education (Nos. 03640270,
05NP0601,
05640325, 05640363).

\newpage
\begin{center}

\end{center}

\newpage
\begin{center}
\section*{Tables}
\end{center}

\begin{table}[h]
\begin{center}
\caption{$s$-wave scattering lengths in units of fm\protect\cite{experiment}.}
\begin{tabular}{lllc}
\\
\hline
$N$-$N$     & ${}^3S_1$  &$-5.432(5)$ \\
            & ${}^1S_0$  &$+20.1(4)$  \\
$\pi$-$N$   & $I=1/2$    &$+0.245(4)$ \\
            & $I=3/2$    &$-0.143(6)$ \\
$\pi$-$\pi$ & $I=0$      &$+0.37(7)$   \\
            & $I=2$      &$-0.040(17)$ \\ \hline
\end{tabular}
\end{center}
\end{table}

\vspace{-1.0cm}

\begin{table}[h]
\begin{center}
\caption{Parameters of simulations. All runs are made at $\beta=5.7$
in quenched QCD.}
\begin{tabular}{lllc}
\\
\hline
            & $K$    &lattice size  &\# conf.  \\ \hline
$N$-$N$     & 0.150  &$20^3\times 20$ &20\\
            & 0.160  &$20^3\times 20$ &30\\
            & 0.164  &$20^3\times 20$ &20\\
$\pi$-$N$   & 0.164  &$12^3\times 20$ &60\\
            & 0.1665 &$12^3\times 20$ &30\\
$\pi$-$\pi$ & 0.164  &$12^3\times 20$ &70\\
\hline
\end{tabular}
\end{center}
\end{table}

\vspace{-1.0cm}

\begin{table}[h]
\begin{center}
\caption{Scattering lengths in lattice units at $\beta=5.7$ in
quenched QCD.  Numbers in square brackets are current algebra predictions
evaluated with the measured values of $m_\pi, m_N$ and $f_\pi$ with the
tadpole-improved $Z$ factor.   }

\begin{tabular}{cccccc}
\\
\hline
$K$      &          &0.150      &0.160     &0.164    &0.1665\\
\hline
$m_\pi$  &          &1.0758(51) &0.6919(63)&0.5081(35)&0.3588(82)    \\
$m_\rho$ &          &1.1302(66)&0.8171(86) &0.7008(67)&0.588(20)\\
$m_N$    &          &1.788(11) &1.302(13) &1.093(20)  &0.951(42)\\
\hline
$N$-$N$  &${}^3S_1$ &$+10.8(1.2)$ &$+9.0(1.6)$ &$+10.8(9)$ &      \\
         &${}^1S_0$ &$+9.2(1.3)$  &$+7.3(1.9)$ &$+8.0(1.1)$&      \\
\hline
$\pi$-$N$&$I=1/2$   &           &          &$+3.04(66)$& --------\\
         &          &           &          &$[+2.59]$  & $[+2.86]$\\
         &$I=3/2$   &           &          &$-1.10(20)$&$-1.31(22)$ \\
         &          &           &          &$[-1.29]$ &$[-1.43]$ \\
\hline
$\pi$-$\pi$&$I=0$   &           &          &$+3.02(17)$&      \\
         &          &           &          &$[+3.32]$  &      \\
         &$I=2$     &           &          &$-0.924(40)$&    \\
         &          &           &          &$[-0.947]$ &     \\
\hline
\end{tabular}

\end{center}
\end{table}

\newpage
\begin{center}
\section*{Figure Captions}
\end{center}


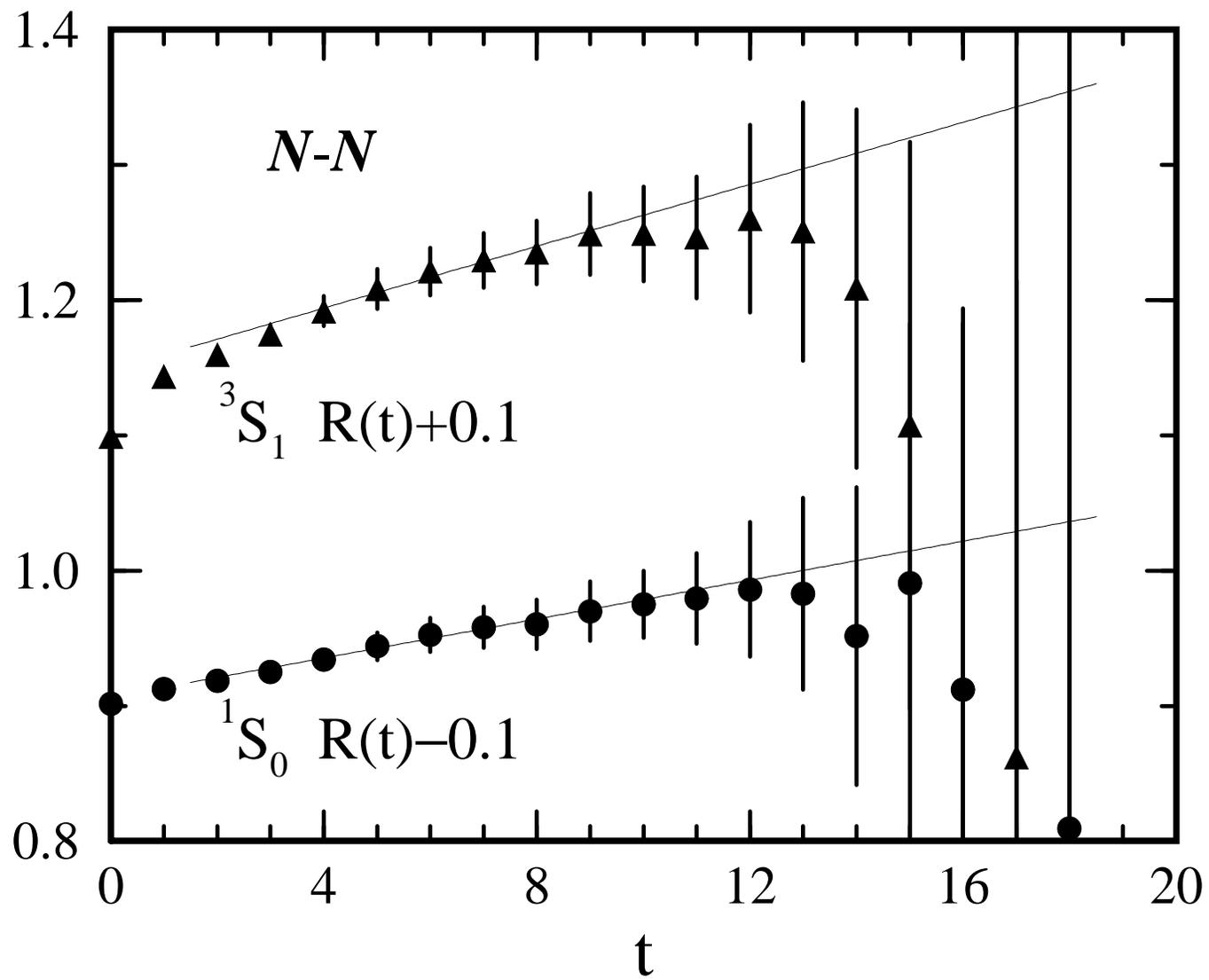
\noindent Fig.~1 : $R(t)$ for the nucleon four-point function for the spin
singlet and triplet channels at $\beta=5.7$ and $K=0.160$ on an $20^3\times
20$
lattice in quenched QCD.  Solid line is the linear fit to  data
over $4\leq t\leq 9$.


\vspace{0.5cm}
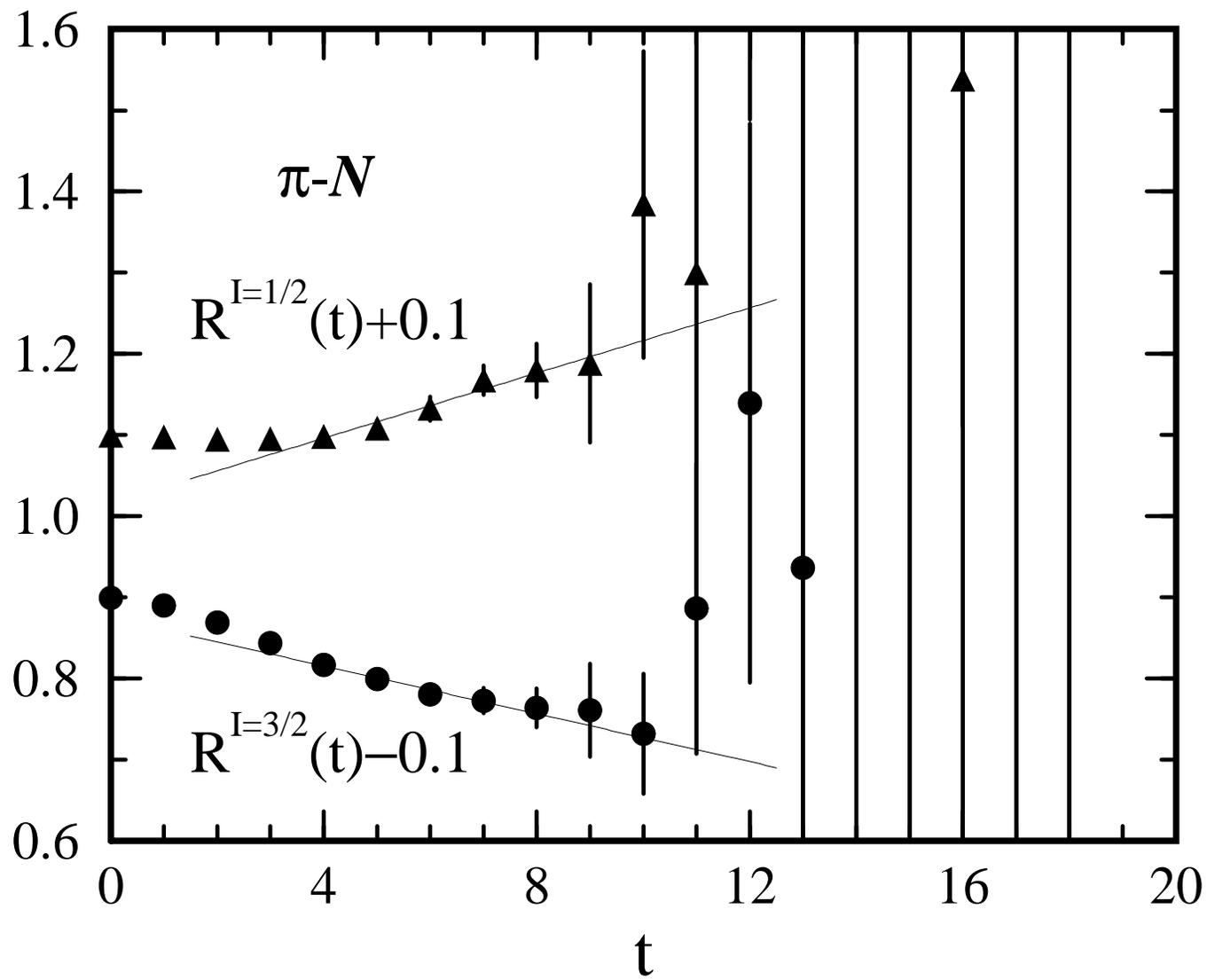
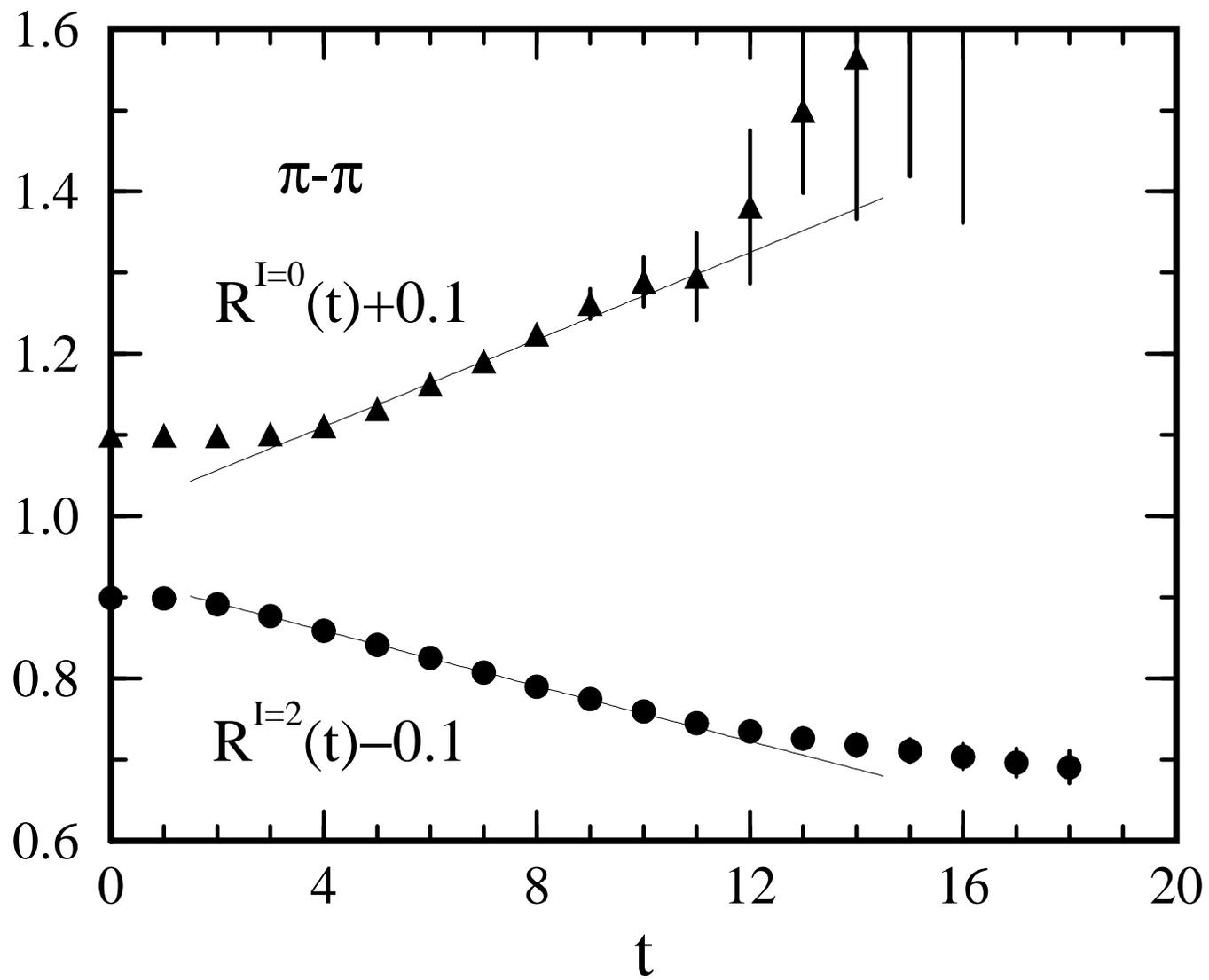
\noindent Fig.~2 : $R(t)$ for the (a) $\pi$-$N$ and (b) $\pi$-$\pi$ four-point
function   at $\beta=5.7$
and $K=0.164$ on an $12^3\times 20$ lattice in quenched QCD.  Solid line is
the
linear fit to  data over $4\leq t\leq 9$.


\vspace{0.5cm}
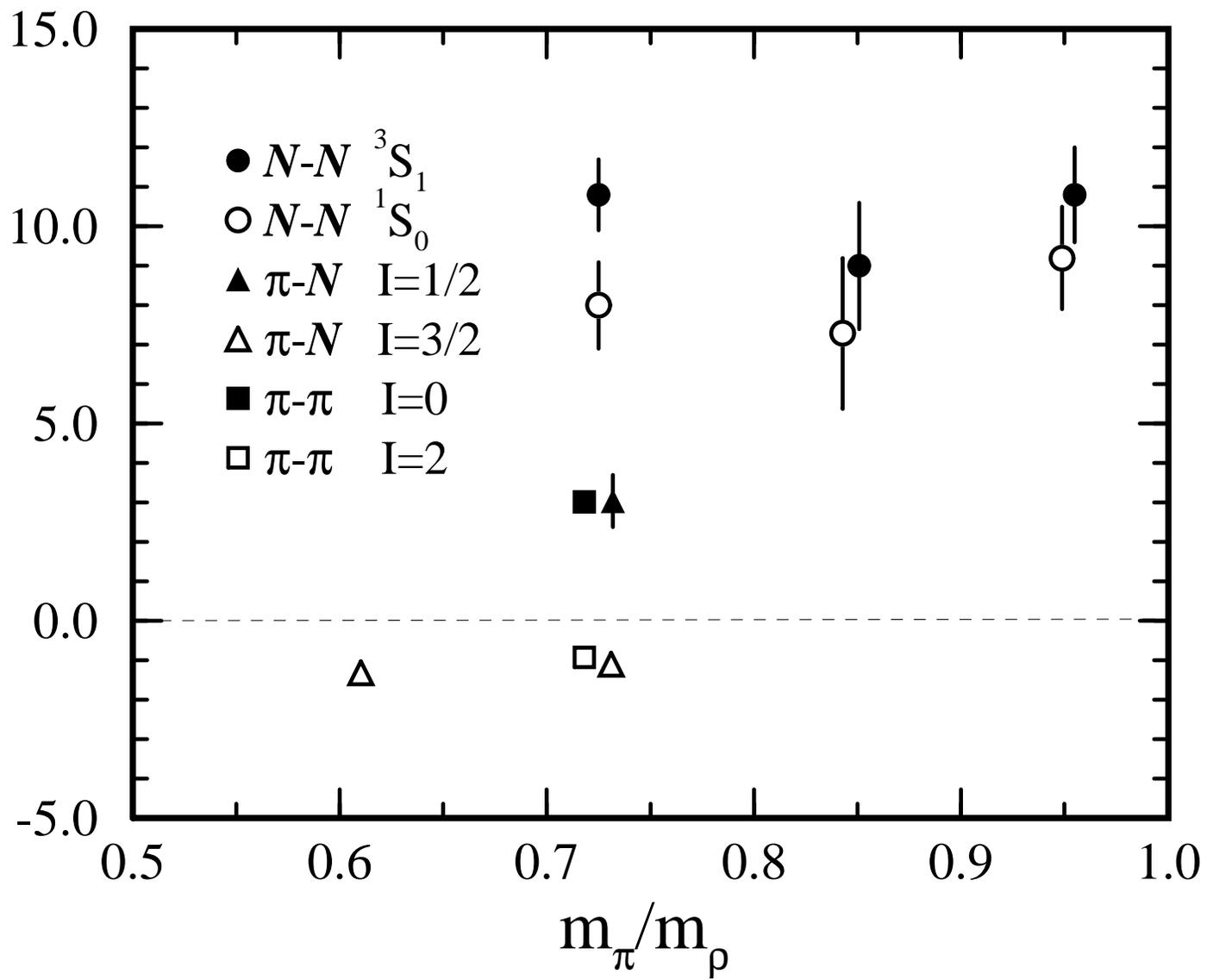
\noindent Fig.~3 : Scattering lengths in lattice units at $\beta=5.7$ in
quenched QCD as a function of $m_\pi/m_\rho$.


\begin{thebibliography}{99}
\bibitem{footnote} This is different from the cases of $\pi$-$\pi$ and
$\pi$-$N$; their dominant low energy force arises from the exchange of a
$\rho$
meson, whose mass, in contrast to the $\pi$ meson mass,  does not differ
radically from the realistic value even for heavy quarks.  For $N$-$N$
scattering an approximate degeneracy of pseudoscalar and vector meson masses
for heavy quarks should lead to a substantial cancellation between the
attractive pseudoscalar exchange and the repulsive vector meson exchange.
\end{thebibliography}
\end{document}